\documentclass[aip, apl, twocolumn, graphicx, groupedaddress, 10pt]{revtex4-1}

\usepackage{graphicx} 
\usepackage{color}

\usepackage[english]{babel}
\usepackage[utf8]{inputenc}

\newcommand{\myUnit}[1]{\,\mathrm{#1}}

\usepackage{amssymb}
\usepackage{hyperref}

\begin{document}
\hyphenation{hetero-struc-ture}

\title[Hybrid Graphene/GaAs nanostructures]{Capacitive coupling in hybrid Graphene/GaAs nanostructures}

\author{Pauline Simonet}
\email{psimonet@phys.ethz.ch}
\author{Clemens R{\"{o}}ssler}
\author{Tobias Kr{\"{a}}henmann}
\author{Anastasia Varlet}
\author{Thomas Ihn}
\author{Klaus Ensslin}
\author{Christian Reichl}
\author{Werner Wegscheider}
\affiliation{Solid State Physics Laboratory, ETH Z{\"{u}}rich, 8093 Z{\"{u}}rich, Switzerland}
\date{\today}

\begin{abstract}
Coupled hybrid nanostructures are demonstrated using the combination of lithographically patterned graphene on top of a two-dimensional electron gas (2DEG) buried in a GaAs/AlGaAs heterostructure. The graphene forms Schottky barriers at the surface of the heterostructure and therefore allows tuning the electronic density of the 2DEG. Conversely, the 2DEG potential can tune the graphene Fermi energy. Graphene-defined quantum point contacts in the 2DEG show half-plateaus of quantized conductance in finite bias spectroscopy and display the 0.7 anomaly for a large range of densities in the constriction, testifying to their good electronic properties. Finally, we demonstrate that the GaAs nanostructure can detect charges in the vicinity of the heterostructure's surface. This confirms the strong coupling of the hybrid device: localized states in the graphene ribbon could in principle be probed by the underlying confined channel. The present hybrid graphene/GaAs nanostructures are promising for the investigation of strong interactions and coherent coupling between the two fundamentally different materials.
\end{abstract}


\maketitle 


The control of individual electrons in semiconductor nanostructures allows the investigation of electron-electron interactions at a basic level \cite{kouwenhoven_few-electron_2001}. Recently, it has become possible to study such phenomena in nanostructured graphene devices \cite{guttinger_charge_2008}. In such experiments, the coupling of neighboring quantum devices has to be understood and controlled in great detail. The combination of electronic devices made from different material systems, such as graphene on GaAs/AlGaAs heterostructures, offers unique opportunities for well-coupled electronic systems with strongly differing energy-momentum relations.

The technological challenge of combining graphene with a GaAs substrate proved to be laborious because of graphene's invisibility on this substrate \cite{calizo_effect_2007,yu_large_2009,wurstbauer_imaging_2010,peters_enhancing_2011,friedemann_graphene_2009}. Using visibility-enhancing GaAs/AlAs superlattices \cite{friedemann_graphene_2009}, Woszczyna {\it et al.} could measure a state-of-the-art quantum resistance standard in a graphene Hall bar on insulating GaAs \cite{woszczyna_magneto-transport_2011, woszczyna_precision_2012}. 
Other works have bypassed the problem of the graphene invisibility by transferring chemical-vapor-deposited graphene on n-type GaAs chips to study the formation of Schottky barriers \cite{tongay_rectification_2012}, and use them to fabricate Schottky-junction based solar cells \cite{jie_graphene/gallium_2013}. 

Fewer works have studied the interaction of large-area graphene with a buried 2DEG in a GaAs/AlGaAs heterostructure. Tang {\it et al.} have demonstrated a micrometer-long graphene field-effect transistor gated by the 2DEG underneath \cite{tang_graphene-gaas/alxga1xas_2012} and a highly tunable GaAs far infrared photodetector covered by a graphene top-gate \cite{tang_highly_2014}. The first Coulomb drag measurements in a micrometer-sized graphene/GaAs 2DEG bilayer system established the importance of Coulomb interactions between both charge carrier systems \cite{gamucci_anomalous_2014}.

In this work, we take the combination of the two materials one step further by forming GaAs nanostructures using Schottky barriers made of graphene. Our first sample is a GaAs quantum point contact (QPC) defined by graphene split-gates and our second sample consists of self-aligned and capacitively coupled graphene/GaAs constrictions. We present the fabrication, functionality, stability and electronic properties of these nanostructures. We finally demonstrate that the GaAs QPC acts as a detector for charges located close to the graphene gates. 


\begin{figure}[t]
\centering
\includegraphics[width=8.5cm]{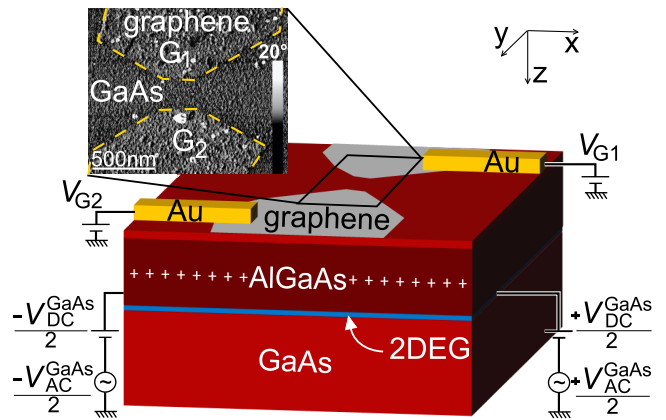}
\caption{Schematic of the first graphene-GaAs hybrid device. The graphene flake (light grey) is clamped on the heterostructure by $50\myUnit{nm}$ thick Ti/Au leads (yellow). The GaAs/AlGaAs heterostructure defines a 2DEG (blue) located $90\myUnit{nm}$ below the surface. The silicon doping is indicated by plus signs. Inset: Atomic force microscopy phase image of the sample surface. The graphene flake (brighter, outlined with orange dashed lines) has been etched to form two split-gates G$_{\rm 1}$ and G$_{\rm 2}$ on the GaAs surface (darker).}
\label{fig1}
\end{figure}

We fabricate hybrid nanostructures with the layer sequence schematically shown in Fig.~\ref{fig1}. The GaAs/Al$_{\rm 0.3}$Ga$_{\rm 0.7}$As heterostructure comprises from top to bottom: a $5\myUnit{nm}$ thick GaAs cap layer on the surface ($z \in [0,5]\myUnit{nm}$), an $85\myUnit{nm}$ thick AlGaAs layer including a Si $\delta-$doping layer of density $7\times 10^{12}\,$cm$^{-2}$ at $z=45\myUnit{nm}$ below the surface, and a heterointerface at $z=90\myUnit{nm}$.

The 2DEG's electron density, measured from the Hall effect, is $n_{\rm{S}}=2.2\,\times 10^{11}\myUnit{cm^{-2}}$ with a mobility of $\mu=3.4\, \times \, 10^{6}\,\rm{cm}^2\rm{V}^{-1}\rm{s}^{-1}$, as measured using the Van der Pauw method without illumination at a temperature of $T=1.3\myUnit{K}$. Graphene flakes are produced by mechanical exfoliation and their monolayer character is verified using Raman spectroscopy \cite{ferrari_raman_2006, graf_spatially_2007}.
After optical lithography steps defining mesa, ohmic contacts and top-gate leads, the graphene flake is transferred onto the GaAs substrate following the method pioneered by Dean {\it et al.} \cite{dean_boron_2010}. Next, an electron beam lithography (EBL) step followed by a Ti/Au  evaporation provides electrical contacts to the flake. The EBL resist is then removed with solvents and using atomic force microscope (AFM) mechanical cleaning \cite{goossens_mechanical_2012}. Finally, a second EBL exposure followed by an O$_{\rm 2}$ plasma ashing step ($200\,$W for $100\,$s) defines the graphene device's shape. The fabrication method and the choice of the etching technique are further explained in the supplemental material \cite{supp_info}. Following these steps, we first fabricated two graphene split-gates, as shown in the inset of Fig.~\ref{fig1}. 

For this device, the differential conductance in the 2DEG $G^{\rm{GaAs}}_{\rm AC}=\mathrm{d}I^{\rm{GaAs}}/\mathrm{d}V^{\rm{GaAs}}_{\rm SD}$ is measured using a standard lock-in technique with a small AC excitation voltage $V^{\rm{GaAs}}_{\rm{AC}}=100\,\mathrm{\mu V~RMS}$ between source (S) and drain (D) contacts at a frequency $f=172.54\,\rm{Hz}$.
All experiments are carried out in a $^4$He cryostat at a temperature of $T\approx 1.4\,\rm{K}$.


\begin{figure}[t]
\centering
\includegraphics[width=8cm]{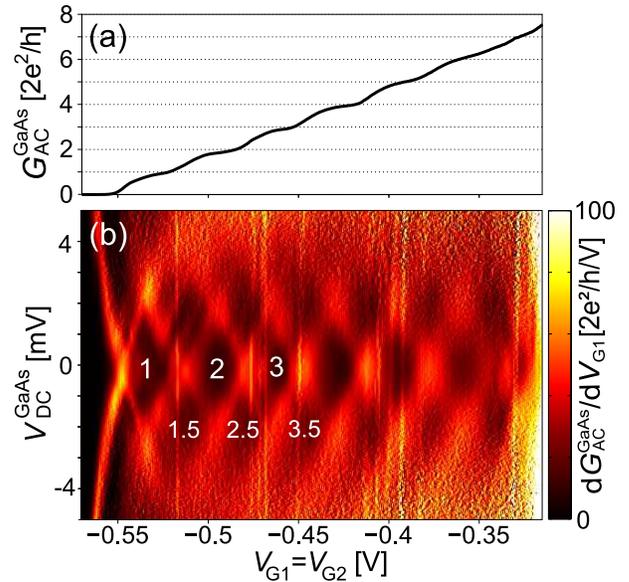}
\caption{(a) GaAs 2DEG differential conductance $G^{\rm GaAs}_{\rm AC}$ measured as a function of split-gate voltages  $V_{\rm G1}=V_{\rm G2}$. (b) Finite bias spectroscopy of the graphene defined QPC. The transconductance $\mathrm{d}G^{\rm GaAs}_{\rm AC}/\mathrm{d}V_{\rm G1}$ is plotted as a function of top-gate voltage ($V_{\rm G1}=V_{\rm G2}$) and the DC bias applied to the 2DEG $V^{\rm GaAs}_{\rm DC}$. Plateaus at integer values of conductance can be seen at low bias and half-plateaus at higher bias. The conductance values for the different plateaus are indicated in $2\,\mathrm{e^2/h}$. For both plots, a serial resistance of $R_\mathrm{S}=2.35\,\mathrm{k}\Omega$ has been subtracted.}
\label{fig2}
\end{figure}

Applying a negative voltage to the two graphene split-gates G$_{\rm 1}$ and G$_{\rm 2}$ depletes the underlying 2DEG. The depletion voltage of graphene and of Ti/Au reference split-gates are the same within our experimental uncertainty (see Section I \cite{supp_info}). Hence, their contact potentials are similar. 
Moreover, we observe quantized conductance until pinch-off is reached, due to the formation of discrete modes in the 2DEG channel between the graphene gates. Figure~\ref{fig2}(a) shows an example of the stepwise decrease in the GaAs differential conductance. As this is a two-point measurement, the resistance of the cables, bond wires, contacts, and mostly of the ohmic contacts to the 2DEG and of the 2DEG itself add to the resistance of the QPC. A serial resistance of $2.35\,\mathrm{k}\Omega$ has thus been subtracted from the data such that the fifth plateau resides at the expected conductance of $5\times 2\,\mathrm{e^2/h}$. 
The observation of six conductance plateaus agrees well with the expectation based on the lithographic gap $w=200\,$-$\,215\,$nm between the graphene split-gates, which corresponds to a number of modes $N\approx w/(\lambda_{\rm F}/2)=2w/\sqrt{2\pi /n_{\rm{S}}}\approx 7\,$-$\,8$. 

In Fig.~\ref{fig2}(b), we show the transconductance $\mathrm{d}G^{\rm GaAs}/\mathrm{d}V_{\rm G1}$, with $V_{\rm G1}=V_{\rm G2}$, as a function of split-gate voltage and DC bias voltage $V^{\rm GaAs}_{\rm DC}$. The above-mentioned modes now draw a diamond pattern. 
Indeed, when increasing the split-gate voltage along $V^{\rm GaAs}_{\rm DC}=0\,$V, each bright peak corresponds to another QPC subband falling below the Fermi level as the confinement potential is lowered. The dark regions correspond to the conductance plateaus occurring when the number of occupied modes is kept constant. As the bias voltage is increased, the bias window opens until it allows one additional subband to contribute to the conduction at the corners of the dark diamonds. Thus, the width in bias voltage of these diamonds gives a QPC subband spacing of $E_{\rm{sub}}\approx 2\,$meV.

Upon further increasing the bias voltage, additional dark regions are observed in Fig.~\ref{fig2}(b). These diamonds are conductance plateaus at half-integer multiples of $2\,\mathrm{e^2/h}$, often called half-plateaus. They occur when a subband-bottom resides between the source and drain chemical potentials \cite{patel_evolution_1991}.  
Half-plateaus can vanish due to scattering events involving the electronic states available at higher bias. Hence, observing them testifies to the cleanliness and electrical stability of the device. Additionally, the quantized conductance plateaus and the half-plateaus have similar sizes, indicating that the confinement potential is close to harmonic \cite{rossler_transport_2011}.
Thus, no observable degradation of transport properties results from the use of graphene top-gates or from the employed processing technology. 


\begin{figure*}[t]
\centering
\includegraphics[width=17cm]{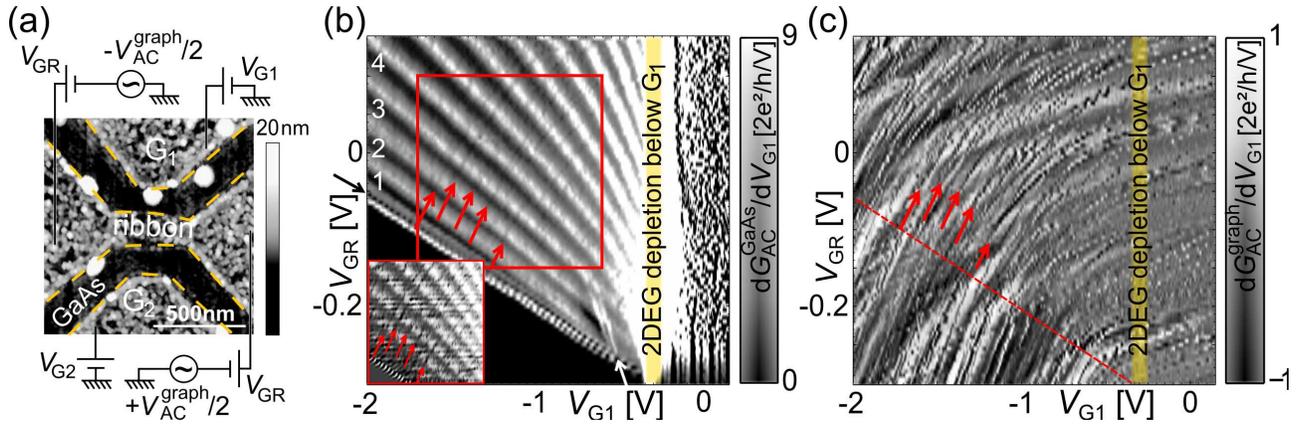}
\caption{(a) AFM topography image of the second sample's surface. A graphene ribbon and its side-gates (brighter, outlined with orange dashed lines) have been patterned on the GaAs surface (darker). They act as two top split-gates G$_{\rm 1}$ and G$_{\rm 2}$ and a central top-gate G$_{\rm R}$ for the GaAs 2DEG. (b) Transconductance of the GaAs 2DEG $\mathrm{d}G^{\rm GaAs}_{\rm AC}/\mathrm{d}V_{\rm G1}$ as a function of $V_{\rm G1}$ and $V_{\rm GR}$ for $V_{\rm G2}=-0.5\,$V. The light yellow band approximately indicates the depletion voltage under G$_{\rm 1}$. The values of quantized conductance in $2\,\mathrm{e^2/h}$ are indicated for the first plateaus. Red arrows mark charge detection events (faint dark lines). A serial resistance of $R_{\rm S}=2.53\,\mathrm{k}\Omega$ has been subtracted from the raw data. Inset: Directional derivative of $G^{\rm GaAs}_{\rm AC}$ along the depletion line of the QPC (data from the red rectangle, interpolated in the $V_{\rm GR}$ direction). (c) Transconductance in the graphene ribbon $\mathrm{d}G^{\rm graph}_{\rm AC}/\mathrm{d}V_{\rm G1}$ as a function of $V_{\rm G1}$ and $V_{\rm GR}$ recorded simultaneously with b. The red dashed line marks the pinch-off voltages of the GaAs QPC recorded in b.}
\label{fig3}
\end{figure*}

Having demonstrated the suitability of graphene as a top-gate, more involved graphene-GaAs hybrid nanostructures have been fabricated. Figure \ref{fig3}(a) shows an AFM topography image of our second device. It consists of a $200\,$nm wide graphene nanoribbon (G$_{\rm R}$) with side-gates G$_{\rm 1}$ and G$_{\rm 2}$ on the same GaAs/AlGaAs he\-tero\-struc\-ture as for the first device. 
The differential conductance in the graphene ribbon is defined like for the 2DEG as $G^{\rm graph}_{\rm{AC}}=\mathrm{d}I^{\rm graph}/\mathrm{d}V^{\rm graph}_{\rm{SD}}$ using the same lock-in measurement technique as previously mentioned, with a source-drain (SD) voltage of $V^{\rm graph}_{\rm{AC}}=V^{\rm GaAs}_{\rm{AC}}=50\,\mathrm{\mu V~RMS}$ for both graphene and 2DEG, and frequencies $f^{\rm{GaAs}}=77.7\,\rm{Hz}$ and $f^{\rm{graph}}=33.3\,\rm{Hz}$. 

The biased side-gates lead to the formation of a QPC in the 2DEG and the ribbon can be used to tune the density in its center. This is depicted in Fig.~\ref{fig3}(b). The QPC transconductance is shown as a function of the gate voltage $V_{\rm G1}$ and the ribbon voltage $V_{\rm GR}$. $V_{\rm G2}$ is kept constant at $-500\,$mV, such that the 2DEG underneath G$_{\rm 2}$ is depleted.
Figure~\ref{fig3}(b) exhibits three regimes of conductance: for $V_{\rm G1}>V_{\rm dep}\approx -0.3\,$V (yellow band), a current can flow below G$_{\rm 1}$ so electrons are not confined in a channel. For more negative $V_{\rm G1}$, a QPC is formed and the conductance decreases stepwise. In this region, a decrease of $V_{\rm GR}$ reduces the electron density in the center of the channel, thus the plateaus and pinch-off positions occur at more positive $V_{\rm G1}$. Finally, below the pinch-off voltage (black region), the QPC is completely depleted and no current can flow.

Below the first plateau at $2\,\mathrm{e^2/h}$, a shoulder at $\approx 0.6\times 2\,\mathrm{e^2/h}$ is observed (black arrow, left of the graph). Its conductance value is stable for a large range of densities, therefore we attribute this shoulder to a quite pronounced 0.7 anomaly. 
Usually observed in clean 2DEGs with moderate bulk electron densities \cite{rossler_transport_2011}, the 0.7 anomaly has been found to be strongest at temperatures around $T\sim 1.5\,\rm{K}$ \cite{thomas_interaction_1998, kristensen_temperature_1998}. In agreement with other studies \cite{thomas_spin_2000, nuttinck_quantum_2000}, the anomaly shifts to lower conductance at low QPC electronic density, reaching $0.4\times 2\,\mathrm{e^2/h}$ for $V_{\rm GR}=-0.26\,$V. Interestingly, we find that this feature survives at lower densities in the channel than the first quantized plateau. This is in agreement with its general robustness with temperature and lateral shifting \cite{thomas_interaction_1998}.

Additional features in the QPC conductance are observed at low $V_{\rm GR}$, i.e. low density in the channel.
A kink occurs on a line crossing the first three plateaus (white arrow). This is most probably the result of impurities creating local minima in the confinement potential of the channel. Electronic bound states form in these minima and transmission resonances appear \cite{weis_transport_1992}.

The current through the graphene ribbon top-gate can be measured simultaneously with the current in the QPC. Fig.~\ref{fig3}(c) shows the graphene transconductance $\mathrm{d}G^{\rm graph}_{\rm AC}/\mathrm{d}V_{\rm G1}$ as a function of $V_{\rm G1}$ and $V_{\rm GR}$ recorded simultaneously with the data shown in Fig.~\ref{fig3}(b). Smooth conductance fluctuations, commonly observed in graphene nanoribbons \cite{han_energy_2007, todd_quantum_2008, minke_phase_2012, bischoff_characterizing_2014}, are visible as a background. They have positive slopes because they occur at a constant values of the ribbon Fermi energy and because $V_{\rm G1}$ and $V_{\rm GR}$ have an opposite influence on them. When $V_{\rm G1}$ decreases, the ribbon Fermi energy increases, while a variation of $|e| V_{\rm GR}$ is directly a variation of its Fermi energy. 

Superimposed onto the broad conductance variations, fine lines (red arrows in the map) also have positive slopes, indicating tunnel-coupling to either the ribbon or G$_{\rm 1}$ (see section IV \cite{supp_info} for more details). However, they probably are not Coulomb resonances occurring in the graphene ribbon. Indeed, the transport gap in the ribbon cannot be reached within the accessible range of side-gates voltages. The graphene conductance in this map is above $0.26\,\mathrm{e^2/h}$ and increases on average with increasing $V_{\rm GR}$ (not shown), suggesting that the flake is $p$-doped.

Instead, these fine lines can be explained by charge detection. Both QPCs and quantum dots have been shown to be sensitive detectors that can resolve a few percents of an electronic charge at a distance of a few hundred nanometers \cite{field_measurements_1993, elzerman_few-electron_2003, rossler_tunable_2013}. The strong conductance fluctuations in graphene nanoribbons make these structures sensitive detectors as well, even outside their transport gap. Variations of charge manifest themselves as kinks in the conductance of the detector, as seen in Fig.~S4(c) \cite{supp_info}. The narrow lines visible in Fig.~3(c) are thus attributed to charge traps detected by the graphene ribbon (see Section III \cite{supp_info} for more details).

Similar measurements on a reference ribbon on SiO$_{\rm 2}$ fabricated in the same way revealed that such resonances may come from residual carbon islands in the etched patterns left by the soft etching \cite{simonet_graphene_2015}. The charge traps could also be impurities implanted during the etching process, but they cannot be located deep in the heterostructure since they should be tunnel-coupled to G$_{\rm 1}$ or the ribbon.

These lines evolve differently in the three regimes mentioned for Fig.~\ref{fig3}(b). For $V_{\rm G1}>V_{\rm dep}\approx -0.3\,$V (yellow band in Fig.~\ref{fig3}(b) and (c), these fluctuations are barely influenced by $V_{\rm G1}$ because of the stronger capacitive coupling between G$_{\rm 1}$ and the 2DEG underneath. Between the depletion voltage $V_{\rm dep}$ below G$_{\rm 1}$ and the pinch-off voltage of the QPC (along the red dashed line in Fig.~\ref{fig3}(c), the electronic density in the GaAs channel is decreased, so the influence of G$_{\rm 1}$ on the charge traps and hence their slopes increase. Finally, the capacitive coupling between G$_{\rm 1}$ and the traps is maximal when the QPC is completely depleted in the pinch-off region.

The same red arrows are drawn in Fig.~\ref{fig3}(b), its inset and Fig.~\ref{fig3}(c). Indicated by these arrows, the faint lines observed in Fig.~\ref{fig3}(c) can also be faintly seen in Fig.~\ref{fig3}(b) and with better visibility in the inset, where a directional derivative has been performed. Thus, the GaAs QPC also detects the same charge traps as seen in the graphene conductance. This means that the system is sufficiently well-coupled to allow the GaAs QPC to detect charges in the graphene plane. The next step could be the fabrication of a similar hybrid graphene-GaAs device with a graphene ribbon whose transport gap is accessible. The transfer of a boron-nitride flake before the graphene flake for instance would permit a more efficient etching process and a smaller doping of the final graphene ribbon. Spatial information on the ribbon's localized charges in the transport gap could be gained using their detection by the GaAs QPC underneath.

In summary, we fabricated capacitively coupled graphene/GaAs nanostructures and characterized them by transport spectroscopy. Conductance quantization in the GaAs QPC could be observed. The observation of finite bias half-plateaus confirmed that this QPC exhibits high purity and charge stability. A second sample including a central graphene top-gate was used to observe the density dependence of the QPC conductance as the 2DEG is depleted. The presence of the 0.7 anomaly for a large range of densities was evidence of the quality of the device. Finally, we demonstrated mutual capacitive coupling between the graphene constriction and the GaAs QPC, including charge detection signals in the QPC conductance.
Further high-quality hybrid nanostructures should allow probing localized states at graphene edges using the QPC defined in the GaAs 2DEG. Using shallower heterostructures, Coulomb drag in hybrid nanostructures or tunneling coupling between quasi-relativistic charge carriers in graphene with massive electrons in GaAs could be investigated.

\begin{acknowledgments}

The authors would like to thank D. Bischoff and A. Kozikov for helpful discussions and R. Gorbachev for his advice on the graphene transfer technique. Support by the Marie Curie Initial Training Action (ITN) Q-NET 264034, the Marie Curie ITN S$^3$ Nano and the ETH FIRST laboratory are gratefully acknowledged.
\end{acknowledgments}

\bibliographystyle{aipnum4-1}
\bibliography{bib_gaas3}

\end{document}